\documentclass[twocolumn]{openjournal}

\makeatletter
\renewcommand\title[2][]{%
  \def\@title{#2}%
  \def\@shorttitle{#1}%
  \let\@AF@join\@title@join
}
\makeatother

\makeatletter

\def\@eapj@cap@font{\bfseries}
\def\@eapj@tabname{Table}
\def\fnum@table{{\@eapj@cap@font \@eapj@tabname~\thetable.---}}

\long\def\@makecaption#1#2{%
  \noindent\begin{minipage}{0.9999\linewidth}%
    \vspace*{\abovecaptionskip}%
    \noindent\footnotesize
    #1 #2\par
    \vskip\belowcaptionskip
  \end{minipage}\par
}

\makeatother

\usepackage{newtxtext,newtxmath}
\usepackage{xcolor}
\usepackage[normalem]{ulem}
\usepackage{fontawesome}
\usepackage{booktabs}
\usepackage{bigdelim}
\usepackage{hyperref}
\hypersetup{
    unicode, 
    colorlinks=true,
    linkcolor=linkcolor,
    citecolor=linkcolor,
    filecolor=linkcolor,
    urlcolor=linkcolor,
}
\usepackage{color,colortbl}
\definecolor{linkcolor}{rgb}{0.0,0.3,0.5}
\usepackage{tensind}
\tensordelimiter{?}
\DeclareGraphicsExtensions{.bmp,.png,.jpg,.pdf}
\usepackage{verbatim}
\usepackage{orcidlink}
\usepackage[T1]{fontenc}

\DeclareRobustCommand{\VAN}[3]{#2}
\let\VANthebibliography\thebibliography
\def\thebibliography{\DeclareRobustCommand{\VAN}[3]{##3}\VANthebibliography}

\usepackage{graphicx}
\usepackage{amsmath}

\begin{document}


\title{gevolution 2.0: GPU-accelerated relativistic $\boldsymbol{N}$-body simulations for cosmology}
\shorttitle{gevolution 2.0}

\author{Julian Adamek \orcidlink{0000-0002-0723-6740}}
\affiliation{Institut f\"ur Teilchen- und Astrophysik, ETH Z\"urich, Wolfgang-Pauli-Strasse 27, 8093 Z\"urich, Switzerland}
\affiliation{Institut f\"ur Astrophysik, Universit\"at Z\"urich, Winterthurerstrasse 190, 8057 Z\"urich, Switzerland}
\affiliation{D\'epartement de Physique Th\'eorique, Universit\'e de Gen\`eve, 24 quai Ernest-Ansermet, 1211~Gen\`eve~4, Switzerland}
\email{adamekj@ethz.ch}

\author{\O{}yvind Christiansen}
\affiliation{CEICO, Institute of Physics of the Czech Academy of Sciences, Na Slovance 1999/2, 182 00, Prague 8, Czechia}

\shortauthors{J. Adamek and \O{}. Christiansen}

\begin{abstract}
High-performance computing is increasingly dominated by hardware acceleration using Graphics Processing Units (GPUs). To take advantage of this long-term trend, we implement a major overhaul of the parallelisation approach in the relativistic particle-mesh $N$-body code \texttt{gevolution}. The new version \texttt{2.0} of \texttt{gevolution} employs three layers of parallelisation: MPI for scalability on a distributed memory system, shared memory parallelisation on each MPI rank using OpenMP, and offloading all compute intensive tasks to GPUs using CUDA. The code also includes many new features that have been developed over the past years. We provide an overview of the code structure and show key performance benchmarks. This publication is accompanied by the simultaneous public release of \texttt{gevolution} \texttt{2.0} and of a GPU-ready version of the \texttt{LATfield2} library which provides the parallelisation backend.
\end{abstract}

\maketitle


\section{Introduction}

General relativity is the foundation of modern cosmology, underpinning our best understanding of gravitational dynamics as well as the projection of observables on our past light cone. To test gravity using the large-scale structure of the Universe, having a fully relativistic forward model can be beneficial: on the one hand, it can predict subtle relativistic effects that are neglected in the Newtonian limit; on the other hand, it provides a natural treatment of ultra-relativistic sources of stress-energy. The latter is particularly relevant for models beyond the current $\Lambda$CDM concordance model, where additional fields could lead to interesting gravitational dynamics that are poorly described by the Newtonian limit.

To implement such a fully relativistic forward model in the context of cosmological simulations, we developed the relativistic particle-mesh $N$-body code \texttt{gevolution}, whose first public release was presented in \citet{Adamek:2015eda,Adamek:2016zes}. Taking advantage of the weak-field conditions that are valid on cosmological scales, \texttt{gevolution} computes the full set of metric perturbations at least to leading weak-field order and places no explicit restrictions on the nature and composition of the stress-energy source. The original code uses MPI for parallelisation and is typically run with one MPI rank per CPU core, as no shared memory parallelisation or GPU support was implemented. In the backend, \texttt{gevolution} uses the \texttt{LATfield2} library \citep{Daverio:2015ryl} to manage the parallelisation, which is scalable beyond the tens of thousands of MPI ranks. The core feature for the gravity solver is the distributed 3D Fast Fourier Transform (FFT) available in \texttt{LATfield2}, allowing us to work in Fourier space for a straightforward scalar-vector-tensor decomposition of the equations and an easy way to solve linear constraints.

Since its initial release, \texttt{gevolution} has been continuously maintained through incremental upgrades. In version \texttt{1.1}, an interface with the linear Einstein-Boltzmann code \texttt{CLASS} \citep{Blas:2011rf} was implemented, providing a powerful mechanism to account for linear perturbations sourced by subdominant components such as radiation and neutrinos \citep{Adamek:2017grt,Adamek:2017uiq}. This version also saw a subtle change in the convention for the scalar metric perturbations, switching from the linear weak-field form, which was due to \citet{Green:2011wc}, to the exponential form, which is more common in the relativistic literature.\footnote{In the notation of \citet{Green:2011wc}, one would, for example, write $g_{00} = -a^2 \left(1 + 2 \Psi\right)$, whereas from version \texttt{1.1} onwards, \texttt{gevolution} uses the convention $g_{00} = -a^2 e^{2\psi}$. At leading order, $\Psi$ and $\psi$ agree, but they differ at higher order. The exponential definition is more convenient for non-perturbative calculations.} The capability to construct light cones on the fly was added in version \texttt{1.2} \citep{Lepori:2020ifz}. Here, \texttt{gevolution} not only produces particle light cones for the construction of halo catalogues but can also store the metric perturbations in a specified spacetime region around the null hypersurface. The latter can then be used for ray tracing, allowing a reconstruction of the fully perturbed light cone in post-processing. As this method uses the metric directly, it is applicable to any geometric theory of gravity and does not rely on any assumption about the relation between matter and gravitational potentials.

Version \texttt{1.2} also forms the base version for several extensions of the code that have been developed as independent branches. Examples include \texttt{k-evolution} \citep{Hassani:2019lmy}, \texttt{asevolution} \citep{Christiansen:2023tfy} and the extension for gravitational wave studies \texttt{AsGRD} \citep{Christiansen:2024uyr}, and recently \texttt{KGB-evolution} \citep{Nouri-Zonoz:2025cul} and \texttt{norns} \citep{Christiansen:2026jnt}. A hybrid implementation that uses \texttt{gevolution} alongside \texttt{Gadget-4} \citep{Springel:2020plp} has been presented by \citet{Quintana-Miranda:2023eyn}.

Several other relativistic codes exist that have been used to simulate the large-scale structure of the Universe; see \citet{Adamek:2020jmr} for an overview and a direct code comparison between \texttt{gevolution} and three other codes. Some of them, like \texttt{CosmoGRaPH} \citep{Giblin:2015vwq} and the \texttt{Einstein} \texttt{Toolkit} \citep{Loffler:2011ay}, are in principle able to probe the strong-field regime, but are often limited in the way they can model the sources of stress-energy. For example, cosmic neutrinos are commonly ignored (\texttt{gevolution} offers several ways to treat them accurately), and the current implementation of the \texttt{Einstein} \texttt{Toolkit} \citep{Macpherson:2016ict} requires a vanishing cosmological constant. The strong-field capabilities are attractive in some contexts, but largely irrelevant for cosmological applications. On the computational side, full numerical relativity is an extremely inefficient way to solve the problem at hand and puts large-volume, high-resolution simulations effectively beyond reach. The key insight here is that linearising the spin-2 field (the only real dynamical degree of freedom of gravity) is an excellent approximation in cosmology that buys essentially all of the performance of traditional simulation codes for large-scale structure without compromising the relativistic footing.

Given that \texttt{gevolution} has a relatively lightweight parallelisation approach, owing to its uniform mesh and the design choice to do without any adaptive refinement, it presents a good use case for hardware acceleration using GPUs. Further motivation can be drawn from the expectation that solving the equations of motion for additional relativistic fields on the mesh, such as the scalar fields introduced in some of the code extensions mentioned above, would typically scale very well on GPUs. Apart from the obvious performance benefits, a GPU implementation also allows utilising the most powerful high-performance computing (HPC) infrastructure available today: as of June 2026, eight of the top ten HPC systems on the TOP500 list\footnote{\url{https://top500.org/lists/top500/2026/06/}} use GPUs, including four out of five existing rated Exascale systems.

In this paper, we present the new GPU implementation of \texttt{gevolution}. In Section\,\ref{sec:method}, we explain the new parallelisation approach and how the code offloads computations to GPUs. In Section\,\ref{sec:equations}, we provide a concise review of the relativistic equations solved by the code. The performance of our new implementation is discussed in Section\,\ref{sec:performance}, and we conclude in Section\,\ref{sec:conclusions}.

\section{Parallelisation Approach}
\label{sec:method}

Using GPUs introduces a number of considerations that should be taken into account when designing a massively scalable code. Current hardware configurations on top-tier HPC systems typically consist of shared-memory nodes with a small number of GPUs and a good number of CPU cores to balance out the performance between CPU and GPU compute capability. The reason is that many applications do not permit complete execution on GPUs, and one wants to avoid a bottleneck on the residual CPU operations. For example, the Nvidia GH200 that we use for our benchmark runs has 72 CPU cores paired with each GPU. While it is possible to share a GPU among multiple MPI ranks, it creates some overhead and may not perform very well for large numbers of ranks. We therefore decided to add shared-memory parallelisation using OpenMP on each rank. In this way, we can bind one MPI rank to each GPU and still use all CPU cores.

Writing and compiling code that can run on GPUs requires a specialised computing stack. There are several competing options available, and we decided to use CUDA. CUDA is a proprietary framework developed by Nvidia, and its main drawback is therefore the fact that it can only be used on Nvidia GPUs. On the other hand, it provides a large ecosystem of optimised libraries and development tools. In addition to being widely adopted in the HPC industry---as of June 2026, Nvidia maintains a share of $87\%$ of the GPU systems on the TOP500 list---Nvidia also completely dominates the consumer market. This means that CUDA code can be developed using consumer hardware: for example, version \texttt{2.0} of \texttt{gevolution} has been tested on a GeForce RTX 3060 that can be purchased for around 300 Swiss francs.

GPUs typically have their own high-bandwidth memory (HBM) used for parallel computations. CUDA provides a unified memory model where CPU and GPU can in principle access any memory location, but for optimal performance, it is good to think about where data should be stored during the simulation. For \texttt{gevolution}, we decided that all particle-mesh operations should be moved to the GPU (device) side, while Fourier representations may reside in CPU (host) memory, i.e.\ regular dynamic random access memory (DRAM). In this way, we try to make optimal use of limited resources: memory is the primary limiting factor of large $N$-body simulations. Bulk memory movement to and from HBM is limited to a small number of places: initialisation and, depending on direction, the final or the initial stage of the 3D FFT, as the Fourier representation is kept on host memory. The packing and extraction of data to and from staging buffers is done on the device side in HBM, and the buffers are moved in bulk to and from DRAM. The same approach is used when particle or pixelised light-cone data are written to disk: output staging buffers are prepared on HBM and then dispatched to DRAM for file output. However, CUDA-aware MPI is used if available, which implies that device memory is used for MPI communication where possible. A breakdown of the memory footprint is shown in Table~\ref{tab:memory}. The code can be run using single or double precision; all benchmarks presented in this paper were run in single precision mode.

\begin{table}
	\centering
    \begin{tabular}{@{}l@{~}lccc}
		\toprule
		 && memory & single & double \\
         && location & precision & precision \\
		\midrule
        &particle data (per particle) & device & 32\,bytes & 56\,bytes\\
        &field component (per grid point) & device & 4\,bytes & 8\,bytes\\
        &Fourier component (per grid point)\! & host & 4\,bytes & 8\,bytes\\
        &FFT workspace (per grid point) & host & 8\,bytes & 16\,bytes\\
        \ldelim\{{3}{*}[\smash{\rotatebox[origin=c]{90}{shared}}] &FFT workspace (per grid point) & device & 12\,bytes & 24\,bytes\\
        &sort workspace (per particle) & device & 36\,bytes & 56\,bytes\\
        &staging buffers & device & variable & variable\\
        &staging buffers & host & variable & variable\\
		\bottomrule
	\end{tabular}
	\caption{Breakdown of the memory footprint. The number of field components varies depending on settings and is typically 12 in real space (stored on device) and 10 in Fourier space (stored on host), using 48\,bytes and 40\,bytes per grid point in single precision, respectively, and twice that number in double precision. Variable staging buffers and other overheads like halo padding typically account for less than $10\%$ of the overall footprint. To conserve device memory, some items use a shared workspace allocation whose size is determined by the largest individual consumer.}
	\label{tab:memory}
\end{table}

\begin{figure}
	\centering
    \includegraphics[width=0.75\columnwidth]{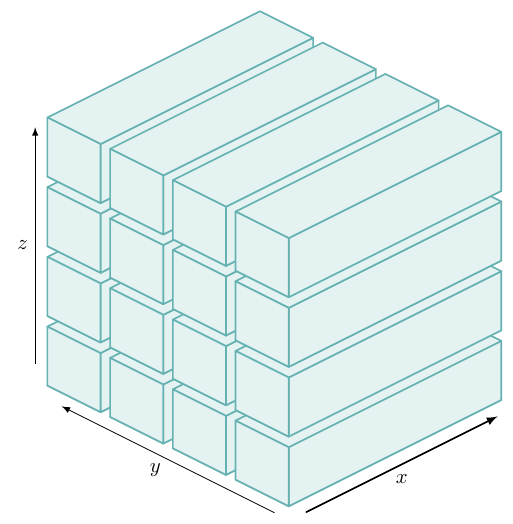}
    \caption{Illustration of the domain decomposition in \texttt{LATfield2}. The `fast' index of the arrays used to store 3D mesh quantities is aligned with the $x$ coordinate.}
    \label{fig:domain}
\end{figure}

In \texttt{gevolution}, the computational domains of the individual MPI ranks are defined by cutting the cubic simulation volume along two of its dimensions in a regular fashion, resulting in rod-shaped domains where the `fast' direction is aligned with the remaining dimension; see Fig.~\ref{fig:domain}. This direction is associated with the $x$ coordinate of the Cartesian coordinate mesh, and by `fast' we refer to the indexing of arrays representing the three-dimensional mesh. This layout is instrumental for the efficient implementation of the distributed 3D FFT, which proceeds by executing a sequence of 1D FFTs and transpositions. Each 1D FFT uses the fact that the entire dimension is local on the task, so no MPI communication is required. Communication takes place at each transposition, which re-arranges a new dimension to become the local one. For details on the original \texttt{LATfield2} implementation, we refer to \citet{Daverio:2015ryl}.

\subsection{The particle handler}

\texttt{LATfield2} was initially developed as a library for mesh-based codes, and its particle handler was added later in the context of the development of \texttt{gevolution}. In its original implementation, the particle handler used doubly-linked lists to manage the particle ensemble. Although easy to implement, this approach risks irregular memory access patterns as the simulation progresses and also adds 16 bytes of memory usage per particle for the two associated list pointers. Later, a leaner version was developed, making half of the pointers redundant by ensuring that the particle lists would only be traversed in one direction. However, this does not remove the constraint that a list can only be traversed in sequence, which makes it a bad design choice for parallelisation. It was therefore decided that the particle handler for \texttt{gevolution} \texttt{2.0} would be rewritten from scratch. This is the single biggest departure from previous versions. However, the new particle handler is written to have an interface very similar to the old one, so that it effectively becomes a drop-in replacement requiring only minimal adjustments.

In CUDA, a kernel launch triggers a large number of parallel threads that are grouped together in thread blocks. Each block consists of a fixed number of threads, e.g.\ 128, which are executed concurrently, whereas the different blocks are executed according to a schedule depending on available GPU resources. Following that logic, we designate each thread block as processing one row of the mesh along the $x$ direction, with the thread blocks then spanning a two-dimensional array over the $y$ and $z$ directions. The particle ensemble itself is not naturally organised into rows, but we can always bin the particles according to which row of mesh cells they fall into and then sort them by their $x$ coordinate. This ordering is not strictly necessary, but it leads to some improved memory access patterns in the particle-mesh interaction. In particular, each thread block will interact only with a small number of rows of the mesh. There are three principal particle-mesh operations.

\textbf{Particle-mesh projection} is a loop over the particles where some quantity, e.g.\ density or stress-energy, is deposited on the mesh. If we use cloud-in-cell (CIC) projection, each thread block will write to a $2\times 2$ set of rows of the mesh. Using sorted particle rows allows us to employ intermediate warp-based reductions to minimise the occurrence of atomic operations. At the boundary of the computational domain, some of the projected quantity needs to be passed on to adjacent MPI ranks. This is done with the help of halo cells that keep track of the spill-over into adjacent domains. After the projection is completed, the contents of these halo cells are communicated via the MPI layer so that they can be added on the other side of the domain boundary. This is the only part of the mesh data that may leave HBM during this operation, although we use CUDA-aware MPI to optimise data movement.

\textbf{Particle kick} is a loop over the particles where their momenta are updated according to their equations of motion. This typically involves metric perturbations and their derivatives, which are defined on the mesh and therefore need to be interpolated to the particle positions. Again, each thread block only needs to access a small number of rows of the mesh, depending on the gradient and interpolation orders. This task is embarrassingly parallel as each particle update is independent. As no MPI communication is required, the particle kick can be executed entirely on the device side, making it by far the fastest component of our particle-mesh algorithm.

\textbf{Particle drift} is a loop over the particles where their positions are updated according to their equations of motion. In simple cases, this may not require a direct interaction with the mesh, but in general relativity the metric plays into this operation. As before, the interpolation of the metric perturbations only touches a small number of rows for each thread block. The drift is also embarrassingly parallel; however, at the end of the update, the particles have to be shuffled back into the rows and ordered. Some particles may have left the computational domain and need to be transferred to adjacent MPI ranks.

To actually store the particle data in HBM, the code allocates three blocks of memory, one each for positions, momenta, and particle IDs. A sorting algorithm sorts the particles according to the row to which they belong and within each row according to their $x$ coordinate. Particles that have left the domain are sorted into eight `special' rows appended at the end. After each drift step, a sort is executed once to fill those `special' rows whose contents are passed through the MPI layer. Then a second sort is required to shuffle in the incoming particles. We use parallel sorting algorithms (\texttt{cub::DeviceRadixSort} and \texttt{cub::DeviceSegmentedSort}) provided by the CUDA Core Compute Libraries. The particle rows are densely packed in memory, and for convenience the code keeps a list of head pointers for the rows that point into the array. Some extra memory is allocated to allow some fluctuation of the particle number inside the computational domain without frequent re-allocation. Ideally, this extra allocation can be chosen at compile time to ensure that no re-allocation is required at all during the run.

In \texttt{LATfield2}, the particle handler takes care of the looping logic and the MPI communication, but the actual operation applied to each particle is left unspecified and can be customised using function pointers. Unfortunately, this is an approach that cannot work on GPUs, as each kernel needs to know at compile time which code it should execute. However, in practice, this code is always known, and using function pointers was just a way to achieve a certain modularity. The same modularity can be achieved with templating, which is the method used in the new particle handler. Instead of accepting a function pointer, the particle-mesh operations can be templated on a callable \texttt{struct} that specifies the code to be executed. This allows the compiler to compile a separate kernel for each type of particle-mesh operation that occurs in the code.

Within \texttt{gevolution}, the particle handler is fitted with input/output methods to allow reading and writing particle data in \texttt{Gadget-2} binary format. The fastest way to dump particle snapshots to disk is to use the \texttt{multi-Gadget2} option, where each MPI rank writes its particles independently to a separate file. As the data reside in HBM, staging buffers are prepared on the GPU side and then moved in bulk to DRAM before writing them to disk.

\subsection{Mesh updates}

The standard MPI-only implementation of \texttt{LATfield2} provides a \texttt{Site} class for indexing the 3D mesh. Whereas simple index operations, such as access to neighbouring sites, can be executed on the GPU side, any \texttt{for} loops using \texttt{Site::next()} would be inherently serial. The new GPU branch of \texttt{LATfield2} therefore provides a generic `\texttt{lattice\_for\_each}' CUDA kernel that can be templated on a callable \texttt{struct} that specifies the code to be executed for each point on the mesh. Similarly to the kernels that deal with particle updates, \texttt{lattice\_for\_each} also provides an easy way to do reductions. Loops over the Fourier mesh are carried out on the CPU side. Here, each loop is parallelised using OpenMP.

\subsection{3D fast Fourier transforms}

The 3D FFT of \texttt{LATfield2} was modified to accommodate the additional layers of parallelisation available in version \texttt{2.0} of \texttt{gevolution}. Most 1D FFTs are executed on device using \texttt{cuFFT}, a CUDA library that provides functionality very similar to \texttt{FFTW3}. However, \texttt{cufftExecR2C}, which is the equivalent to \texttt{FFTW3}'s \texttt{fftwf\_execute\_dft\_r2c}, has an additional requirement on the alignment of the first element in the source array. Due to halo cells and the existence of multi-component fields such as vectors and tensors that need to be Fourier transformed, the required alignment of the memory address cannot be guaranteed. Instead of copying the data into a fresh array, which would be a possible workaround, the code checks the alignment of the data and falls back to \texttt{FFTW3}'s \texttt{fftwf\_execute\_dft\_r2c} for the 1D FFTs where the requirement set by \texttt{cuFFT} is not fulfilled.

As the 3D FFT is decomposed into a sequence of 1D FFTs which use temporary memory for intermediate results, only the first 1D FFT in the forward transform and the last 1D FFT in the backward transform need special memory handling; all other 1D FFTs can be performed on device using \texttt{cuFFT}. Furthermore, in the case of multi-component fields, the alignment meets the requirement for every other component.

Between the 1D FFTs, the data have to pass through the MPI layer for the distributed transpositions. The GPU would be mostly idle during that time, especially when data need to be routed through the inter-node network. However, for multi-component fields, the code does something slightly more optimal: it processes two components at a time concurrently, and offloads the 1D FFTs for each component whenever the other component needs to do a transposition. As a result of this optimisation, the 3D FFT is in practice dominated by the cost of MPI communication.

\newpage
\section{System of equations}
\label{sec:equations}

The gravity solvers of \texttt{gevolution} are based on a weak-field description of general relativity, taking advantage of the helpful behaviour of the Poisson gauge in cosmological settings. The details of the description have evolved somewhat over the years, so we take this opportunity to review the current state of affairs.

\subsection{The relativistic equations}

The line element of the cosmological spacetime is written in the familiar 3+1 decomposition,
\begin{equation}
    ds^2 = g_{\mu\nu} dx^\mu dx^\nu = -\alpha^2 d\tau^2 + \gamma_{ij} \left(dx^i + \beta^i d\tau\right) \left(dx^j + \beta^j d\tau\right)\,,
\end{equation}
where $\alpha$ is the lapse function, $\beta^i$ is the shift vector, and $\gamma_{ij}$ is the induced metric on the spacelike hypersurfaces of constant coordinate time $\tau$. In the Poisson gauge, we choose the parametrisation
\begin{eqnarray}
    \alpha &=& a e^\psi\,,\\
    \gamma_{ij} &=& a^2 e^{-2\phi} \left(\delta_{ij} + h_{ij}\right)\,,
\end{eqnarray}
where $a(\tau)$ is a time-dependent conformal factor (the scale factor), $\psi$ and $\phi$ are the two scalar gravitational potentials and $h_{ij}$ is a spin-2 field that describes gravitational waves. In the Poisson gauge, we further impose that the shift vector $\beta^i$ has a vanishing divergence and, therefore, can be identified as the frame-dragging potential. In the notation used in most of the literature on \texttt{gevolution}, the shift is parameterised as $\gamma_{ij} \beta^j = -a^2 B_i$. The spatial three-metric $\gamma_{ij}$ is written so that $x^i$ are the usual Cartesian comoving coordinates.

To obtain a tractable set of equations for the metric perturbations $\psi$, $\phi$, $B_i$ and $h_{ij}$, weak-field conditions are crucial. In particular, it is assumed that the equations can be linearised in the spin-2 amplitude $h_{ij}$, which is certainly an extremely good approximation in cosmology. This assumption effectively decouples the other gravitational fields from $h_{ij}$ so that they are now given by straightforward constraints. It is worth pointing out that this approximation alone relaxes the stepping criterion of the time integrator typically by two to three orders of magnitude, making it the key step for efficient cosmological simulations.

We reduce complexity even further by also linearising in the frame-dragging potential $B_i$. However, $\psi$ and $\phi$ should not be linearised, as in the absence of sufficiently strong vector and tensor sources in the matter sector some quadratic terms in $\phi$, $\psi$ are relevant for an accurate prediction of $B_i$ and $h_{ij}$. In the weak-field expansion employed in \citet{Adamek:2013wja,Adamek:2016zes} these were identified as the terms with two spatial derivatives. Although this reasoning is sound, there is an additional benefit in keeping all second-order terms in the potentials: by doing so, the numerical scheme incorporates the full second-order relativistic perturbation theory. This means that even on scales close to the cosmological horizon, where relativistic corrections to the Newtonian limit are most relevant, the scheme is equipped to make accurate predictions for the three-point function. For further details, see \citet{Adamek:2021rot,Montandon:2022ulz}.

After neglecting higher-order contributions from $B_i$ and $h_{ij}$, the full Hamiltonian constraint reads
\begin{equation}\label{eq:fullHc}
    e^{2\phi} \Delta\phi - \frac{1}{2} e^{2\phi} \left(\nabla\phi\right)^2 + \frac{3}{2} \left(\mathcal{H} - \partial_\tau\phi\right)^2 e^{-2\psi} = 4 \uppi G a^4 e^{2\psi} T^{00}\,,
\end{equation}
where $\mathcal{H} = \partial_\tau \ln a$ and the spatial gradients are partial derivatives with respect to the coordinates, not covariant ones. The left-hand side of Eq.~\eqref{eq:fullHc} neatly highlights the two contributions to the spacetime curvature: the first two terms are the three-curvature scalar on the constant-time hypersurface, whereas the third term originates from the extrinsic curvature associated with the foliation. The latter also elucidates the role of the conformal factor $a$ that could, in principle, be absorbed in $\psi$, $\phi$ by setting $\psi \rightarrow \psi + \ln a$, $\phi \rightarrow \phi - \ln a$. Beyond perturbation theory, there is no unique equation for $a$, but from a numerical point of view it is useful to introduce it to absorb a large and spatially smooth contribution to the extrinsic curvature. We stress that this does not introduce a background dependence for observables: the functional form of the scale factor is chosen for computational convenience, and the spatially homogeneous modes of $\phi$ and $\psi$ adjust themselves accordingly \citep{Adamek:2013wja,Adamek:2016zes}.

To compute the full set of metric perturbations, we can use the trace-free part of the space-space Einstein equations, which after linearising in $B_i$ and $h_{ij}$ reads
\begin{multline}
    \partial_\tau^2 h_{ij} + 2 \mathcal{H} \partial_\tau h_{ij} - \Delta h_{ij} + \left(\partial_\tau + 2 \mathcal{H}\right) \left(\nabla_i B_j + \nabla_j B_i\right)\\
    + 2\left(\delta_i^k \delta_j^l - \frac{1}{3} \delta_{ij} \delta^{kl}\right) \left[\nabla_k\nabla_l \left(\phi - \psi\right) + 2 \nabla_k\phi \nabla_l\phi\right.\\
    \left.- \left(\nabla_k\phi + \nabla_k\psi\right)\left(\nabla_l\phi + \nabla_l\psi\right)\right] = 16 \uppi G \left(T_{ij} - \frac{1}{3} \delta_{ij} \delta^{kl} T_{kl}\right)\,.\label{eq:spacespace}
\end{multline}
Moving the quadratic terms in $\nabla\phi$, $\nabla\psi$ to the right-hand side and treating them as part of the source, we obtain a clean separation into scalar ($\chi = \phi - \psi$), vector ($B_i$) and tensor ($h_{ij}$) sector. The code explicitly computes the gravitational slip $\chi$ and keeps no separate copy of $\psi$; where required, $\psi$ is computed inline as $\psi = \phi - \chi$. In practice, we compute the nonlinear source term in configuration space and then move to Fourier space where the scalar-vector-tensor decomposition is given by simple projection operators. We can also note that the second-order contributions from $\phi$ and $\psi$, which we treat as part of the source, are only really relevant when the anisotropic stress of the matter is itself of similar amplitude or smaller. For models with large anisotropic stress, these extra contributions are subdominant and could be neglected entirely.

The frame-dragging potential $B_i$ can also be computed from the momentum constraint,
\begin{equation}\label{eq:momentumcontraint}
    -\frac{1}{4}\Delta B_i - \nabla_i \partial_\tau \phi - \left(\mathcal{H} - \partial_\tau\phi\right) \nabla_i \psi = 4 \uppi G a^2 e^{2\psi} T^0_i\,,
\end{equation}
which gives a simple algebraic expression in Fourier space once the spin-1 projector is applied. The solutions of $B_i$ from Eqs.~\eqref{eq:spacespace} and \eqref{eq:momentumcontraint} are not independent, of course, and their equivalence (partially) expresses the covariant conservation of stress-energy. The code allows one to use either equation in the gravity solver: the elliptic constraint \eqref{eq:momentumcontraint} is more robust against strong time evolution of the stress-energy that can occur in exotic models beyond $\Lambda$CDM, but it requires the computation of the momentum density and its Fourier transform.

The constraints for $\phi$ and $\psi$ from Eqs.~\eqref{eq:fullHc} and \eqref{eq:spacespace} remain coupled and could be solved via a predictor-corrector loop. Such a loop works by collecting all the linear terms in $\phi$ and $\psi$ and treating everything else as the source of a set of coupled linear partial differential equations. The nonlinear contributions of $\phi$ and $\psi$ to the source are computed from an initial guess and then repeatedly updated with the solution of the linear system until the solution converges. Here, weak-field conditions control the rate of convergence of the fixed-point iteration. In practice, the evolution of $\phi$ and $\psi$ is sufficiently slow that the initial guess can be obtained from the previous time step and no actual loop iterations are required. The code therefore effectively uses a lagged first-order time-integration of these nonlinear weak-field terms.

At this point, it is also worth commenting that most of the previous literature on \texttt{gevolution} presents the Hamiltonian constraint \eqref{eq:fullHc} in a somewhat different form. Noting that
\begin{equation}
    4 \uppi G a^4 e^{2\psi} T^{00} = -4 \uppi G a^2 T^0_0 + 4 \uppi G a^2 \beta^i T^0_i\,,
\end{equation}
the right-hand side of Eq.~\eqref{eq:fullHc} can be written in terms of the mixed-index quantity $T^0_0$ and a term that---by virtue of the momentum constraint \eqref{eq:momentumcontraint}---is neglected when performing the weak-field truncation. In other words, the previous formulation based on $T^0_0$ and the formulation shown in Eq.~\eqref{eq:fullHc} are both equally valid interpretations in the current approximation scheme. However, moving forward, we think that the form presented here is more useful beyond the specific weak-field truncation of the original implementation.

The $N$-body ensemble is evolved using the geodesic equation for massive particles, which can be obtained the most easily from the general relativistic one-particle Hamiltonian
\begin{equation}\label{eq:1pclH}
    H_\mathrm{p} = \alpha \sqrt{m^2 + \gamma^{ij} q_i q_j} - \beta^k q_k\,,
\end{equation}
where $q_i$ is the canonical momentum and $m$ is the rest mass of the particle. It is worth noting that Eq.~\eqref{eq:1pclH} is exact, giving rise to the Hamiltonian equations of motion
\begin{equation}
    \frac{\partial x^i_\mathrm{p}}{\partial\tau} = \frac{\alpha \gamma^{ij} q_j}{\sqrt{m^2 + \gamma^{kl} q_k q_l}} - \beta^i\,,\label{eq:drift}
\end{equation}
and
\begin{equation}
    \frac{\partial q_i}{\partial\tau} = -\sqrt{m^2 + \gamma^{jk}q_j q_k} \nabla_i \alpha + q_j \nabla_i \beta^j - \frac{\alpha q_j q_k \nabla_i \gamma^{jk}}{2 \sqrt{m^2 + \gamma^{lm} q_l q_m}}\,,\label{eq:kick}
\end{equation}
which specify the drift and kick operation, respectively, for the particle evolution. In \texttt{gevolution}, these operations are applied in a staggered leap-frog scheme, where we expand for small $\phi$, $\psi$ and neglect the contribution of $h_{ij}$. The scattering of particles on gravitational waves is typically so small that it would be lost in numerical round-off errors, so we avoid the additional operations needed to include them in the particle update. Note, however, that we do not assume small momenta, so that our integration scheme remains valid for relativistic particles. This is useful, for example, in the context of simulating massive neutrinos \citep{Adamek:2017uiq,Euclid:2022qde}.

The canonical momentum also allows one to write the stress-energy tensor of the particle ensemble in a compact form,
\begin{eqnarray}
    T^{00}&=&\sum_\mathrm{p} \delta^{(3)}_\mathrm{D}(\boldsymbol{x}-\boldsymbol{x}_\mathrm{p}) \frac{\sqrt{m^2 + \gamma^{ij}q_i q_j}}{\alpha^2 \sqrt{\gamma}}\,,\\
    T^0_i&=&\sum_\mathrm{p} \delta^{(3)}_\mathrm{D}(\boldsymbol{x}-\boldsymbol{x}_\mathrm{p}) \frac{q_i}{\alpha \sqrt{\gamma}}\,,\\
    T_{ij}&=&\sum_\mathrm{p} \delta^{(3)}_\mathrm{D}(\boldsymbol{x}-\boldsymbol{x}_\mathrm{p}) \frac{q_i q_j}{\sqrt{\gamma}\sqrt{m^2 + \gamma^{kl} q_k q_l}}\,,\label{eq:matterstress}
\end{eqnarray}
where $\gamma = a^6 e^{-6\phi}$ is the determinant of the three-metric, and the sum runs over all particles with positions denoted $\boldsymbol{x}_\mathrm{p}$. In a particle-mesh scheme like the one used in \texttt{gevolution}, the stress-energy tensor is projected onto the mesh by convolving the Dirac delta functions with a projection kernel. We use cloud-in-cell projection in most cases, but some projections can use a mix of cloud-in-cell and nearest grid-point assignment. As in the particle update scheme, we keep the momentum non-perturbative and apply the corrections due to metric perturbations consistently in a weak-field expansion.

\subsection{The Newtonian limit}

The Newtonian limit has been studied since the inception of general relativity, including by \citet{Einstein:1916vd} himself. In typical non-cosmological settings with Minkowski-space asymptotics ($\mathcal{H} \rightarrow 0$), the limit is usually taken by neglecting vector and tensor metric perturbations and further assuming non-relativistic matter with $m^2 \gg \gamma^{ij} q_i q_j$. In this case, the Poisson gauge reduces to the Newtonian gauge, and if the linearised gravitational field is sourced by non-relativistic matter only, we have $\phi \simeq \psi$ from Eqs.~\eqref{eq:spacespace} and \eqref{eq:matterstress}. The Hamilton\-ian constraint \eqref{eq:fullHc} reduces to $\Delta \psi \simeq 4 \uppi G \rho$ and the particle equations of motion \eqref{eq:drift} and \eqref{eq:kick} become $\partial x^i_\mathrm{p} / \partial \tau \simeq \delta^{ij} q_i / m$ and $\partial q_i / \partial\tau \simeq - m \nabla_i \psi$, respectively. The latter two imply the familiar Newtonian force law.

The Newtonian limit is somewhat more subtle in cosmological settings ($\mathcal{H} \neq 0$) which have non-vacuum asymptotics. After subtracting the Friedmann equation from the Hamiltonian constraint~\eqref{eq:fullHc}, which removes a large and spatially uniform component of the extrinsic curvature, we are left with
\begin{equation}\label{eq:NewtonHc}
    \Delta \phi - 3 \mathcal{H} \partial_\tau \phi - 3 \mathcal{H}^2 \psi = 4 \uppi G a^2 \delta \rho\,,
\end{equation}
where
\begin{equation}\label{eq:Newtondens}
    \delta \rho = \left(\sum_\mathrm{p} \delta^{(3)}_\mathrm{D}(\boldsymbol{x}-\boldsymbol{x}_\mathrm{p}) \frac{m}{a^3} \left(1 + 3 \phi\right)\right) - \bar{\rho}\,.
\end{equation}
Here, $\bar{\rho}$ is the uniform density that determines the value of $\mathcal{H}^2$ via the Friedmann equation, or vice versa. To obtain a Poisson equation that is more similar to the non-cosmological case, one would like to neglect all occurrences of $\phi$ and $\psi$ in Eqs.~\eqref{eq:NewtonHc} and \eqref{eq:Newtondens} except for the $\Delta\phi$-term, which we may then replace by $\Delta\psi$ due to the fact that $\phi \simeq \psi$ still holds. This is justified under the additional assumption that we are deep inside the horizon, so that $\Delta\phi \gg \mathcal{H} \partial_\tau \phi + \mathcal{H}^2 \psi$. This reduces Eq.~\eqref{eq:NewtonHc} to the standard Poisson equation and also implies\footnote{This can be seen from dimensional analysis: $\Delta \phi \sim L^{-2} \phi$, where $L$ is a typical comoving length scale; furthermore, $4 \uppi G a^2 \delta\rho \sim \mathcal{H}^2 \delta\rho/\bar{\rho}$ due to the Friedmann equation; the statement then follows from $L \mathcal{H} \ll 1$ (sub-horizon scale).} that $\delta\rho/\bar{\rho} \gg \phi$, which guarantees that we can compute the right-hand side of Eq.~\eqref{eq:NewtonHc} from simple count-in-cell densities without considering the volume perturbation in Eq.~\eqref{eq:Newtondens}.

In the expanding Universe, the particle equations of motion \eqref{eq:drift} and \eqref{eq:kick}  for non-relativisitc matter reduce to $\partial x^i_\mathrm{p} / \partial \tau \simeq a^{-1} \delta^{ij} q_i / m$ and $\partial q_i / \partial\tau \simeq - a m \nabla_i \psi$. The acceleration equation therefore reads $\partial^2x^i_\mathrm{p} / \partial\tau^2 \simeq -\mathcal{H} \partial x^i_\mathrm{p} / \partial\tau - \delta^{ij} \nabla_j \psi$, which contains the usual damping term due to expansion.

Because \texttt{gevolution} explicitly runs in the Poisson gauge and keeps track of metric contributions that become relevant at the scale of the horizon and beyond, it offers a relativistic framework that remains valid on all scales that satisfy weak-field gravity conditions. However, it has been shown by \citet{Fidler:2015npa} that the standard Newtonian equations can also be recovered on those scales if one adopts a different gauge choice. The crucial insight of their approach is that the two scalar gauge modes provide sufficient freedom to render all linear weak-field terms into the Newtonian form. Although a very elegant result, it does not generalise beyond the linear weak-field order and comes at the cost of an unusual gauge choice that can muddy the interpretation of simulation results. Nevertheless, in this way it is at least possible to obtain a robust relativistic interpretation of horizon-scale Newtonian simulations at all. In \texttt{gevolution}, the Newtonian limit is therefore implemented by switching explicitly to the so-called $N$-body gauge, which is one of the gauge choices that lead to Newtonian equations in the weak-field limit.

Specifically, the $N$-body gauge can be defined through
\begin{eqnarray}
    \alpha &=& a\,,\\
    \gamma_{ij} \beta^i &=& -a^2 \nabla_i B\,,\\
    \gamma_{ij} &=& a^2 \left[\delta_{ij} - 2\left(\nabla_i\nabla_j - \frac{1}{3} \delta_{ij} \Delta\right) H_\mathrm{T}\right]\,,
\end{eqnarray}
where $H_\mathrm{T}$ is a nearly time-independent large-scale potential that can be predicted from linear perturbation theory, and the potential $B$ is related to the Newtonian gravitational potential $\psi_\mathrm{N}$ via the Hamiltonian constraint as $\psi_\mathrm{N} = - \mathcal{H} B - \partial_\tau B$. Note that $H_\mathrm{T}$ does not appear in the equations of motion of non-relativistic matter and therefore only becomes relevant for mapping simulation output to observables on the light cone; see \citet{Fidler:2017pnb,Adamek:2017kir,Adamek:2019aad} for more details. In the $N$-body gauge, all scalar metric perturbations have been moved to the off-diagonal components of the metric.

One of the advantages of the $N$-body gauge is that it allows for the consistent treatment of species that do not strictly fit into the Newtonian approximation. Their effect on the equations of motion is well-defined in the $N$-body gauge and can be computed, for example, from linear theory where appropriate. In \texttt{gevolution}, such a treatment is implemented for radiation, neutrinos, and linear perturbations of a generic fluid that could model dynamical dark energy. It is worth noting that one can construct other `Newtonian motion' gauges that may absorb such effects directly into the gauge condition \citep{Fidler:2017pnb,Partmann:2020qzb,Fidler:2026wqg}.

\section{Code Performance}
\label{sec:performance}

Performance was measured on the \textit{`Alps'} HPC platform at the Swiss National Supercomputing Centre (CSCS), an HPE Cray EX254n with four Nvidia GH200 (Grace Hopper) Superchips per node, i.e.\ four GPUs and $4\times72$ CPU cores. Our default parallel configuration is therefore four MPI ranks per node, with each rank utilising one GPU and $72$ OpenMP threads (one per CPU core).

\subsection{Strong scaling performance}

\begin{figure}
	\includegraphics[width=\columnwidth]{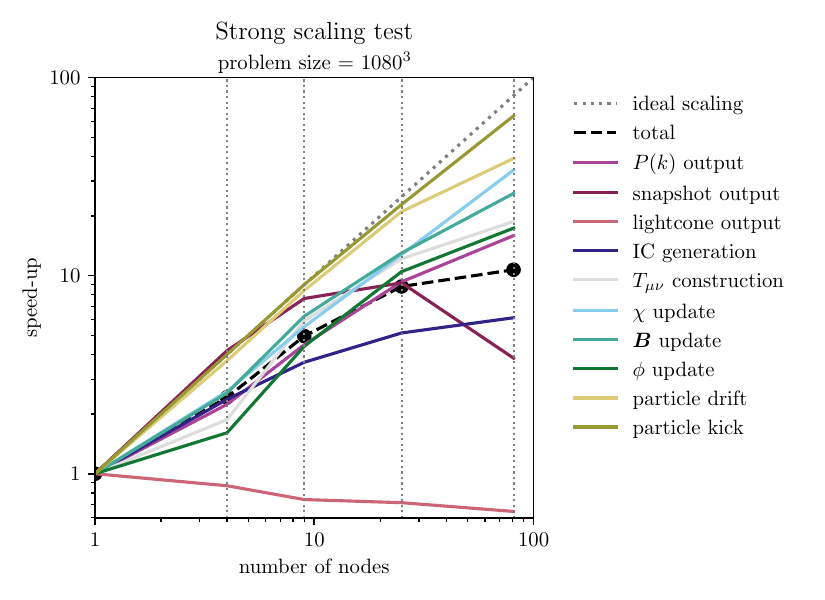}
    \caption{Results of a strong scaling test. The black dashed line indicates the overall performance for a simulation with $1080^3$ particles on a $1080^3$ mesh. The different coloured lines show the scaling behaviour of individual code components as indicated in the legend. Scaling is mainly limited by I/O: on $81$ nodes, the simulation runs in $\sim 93\,\mathrm{s}$ while producing $400\,\mathrm{GB}$ of data that are mostly written using MPI collectives involving all $324$ MPI ranks.}
    \label{fig:strongscaling}
\end{figure}

Strong scaling measures parallel efficiency as one increases the number of processing elements at fixed problem size. Ideal scaling is said to be achieved when the time to solution is inversely proportional to the number of processing elements. Fig.~\ref{fig:strongscaling} shows the performance of \texttt{gevolution} \texttt{2.0} for a strong scaling test using a simulation with $1080^3$ particles and a mesh with the same number of cells. Here, the time to solution is reported as speed-up relative to a run using a single node. The overall performance, shown as black dashed line, can be broken down into different parts, shown in colours. We note that particle updates (drift and kick) achieve near perfect scaling, and that the parts that rely on the distributed 3D FFT also follow an excellent power law, although the growing communication overhead carries a noticeable cost. Limited scaling can be observed in the snapshot output to disk, mainly caused by saturation of the I/O throughput using the HDF5 parallel library. The output of lightcone data, on the other hand, shows almost no speed-up at all and therefore is mainly responsible for the levelling off of the total performance at high node count.

\begin{table}
	\centering
    \begin{tabular}{cccccccc}
		\toprule
		 & total & $\phi$ & $\boldsymbol{B}$ & $\chi$ & kick & drift & $T_{\mu\nu}$ \\
		\midrule
        time (ms) & 2104 & 295 & 455 & 1140 & 12 & 31 & 171\\
		\bottomrule
	\end{tabular}
	\caption{Breakdown of one integration step of a typical run with $1080^3$ particles and a $1080^3$ mesh. The run was executed on 4 nodes using 16 MPI ranks. The breakdown shows how much time is spent (on average) on different tasks during one integration step. The items $\phi$, $\boldsymbol{B}$, $\chi$ refer to the solvers for the respective metric components, which are all based on 3D FFTs. Kick and drift refer to the update of particle momenta and positions, respectively. The item $T_{\mu\nu}$ includes particle-mesh projections for the stress-energy tensor but is dominated by the CLASS interface to model linear contributions (radiation and neutrinos), which also requires 3D FFTs.}
	\label{tab:breakdown}
\end{table}

Generally speaking, due to the high performance of the code on computations, scaling depends quite heavily on the chosen I/O profile, with runs able to scale significantly better if they write fewer data. Here, we want to test all the components of the code and, therefore, our benchmark utilises all the features of the code in a meaningful way. In particular, the simulation used for the strong-scaling analysis writes six snapshots of the particles and the full metric, a deep light cone covering $1000\,\mathrm{deg}^2$, a total of $121$ on-the-fly power spectra for various quantities, and uses the \texttt{CLASS} interface to account for the effect of massive neutrinos in a minimal-mass scenario. The total data volume written to disk amounts to $400\,\mathrm{GB}$.

Table~\ref{tab:breakdown} shows the breakdown of the work performed in one integration step of a representative run of the strong-scaling analysis. The timings are strongly dominated by the distributed 3D FFT, which takes about $148\,\mathrm{ms}$ per field component in this example. Solving for $\phi$ requires two FFTs (one forward and one backward), $\chi$ requires seven (six forward and one backward), $\boldsymbol{B}$ requires three (backward only, reusing the same Fourier source as $\chi$), and the construction of $T_{\mu\nu}$ requires one backward transform for the linear species (radiation and neutrinos). Therefore, of the $2.1\,\mathrm{s}$ taken for one integration step, the code spends about $1.9\,\mathrm{s}$ on FFTs, while the kick, drift, and particle-mesh projections together account for less than $10\,\%$ of the execution time. Within the FFTs, the code spends most of the time on MPI communication for the required transpositions.

\subsection{Weak scaling performance}

\begin{figure}
	\includegraphics[width=\columnwidth]{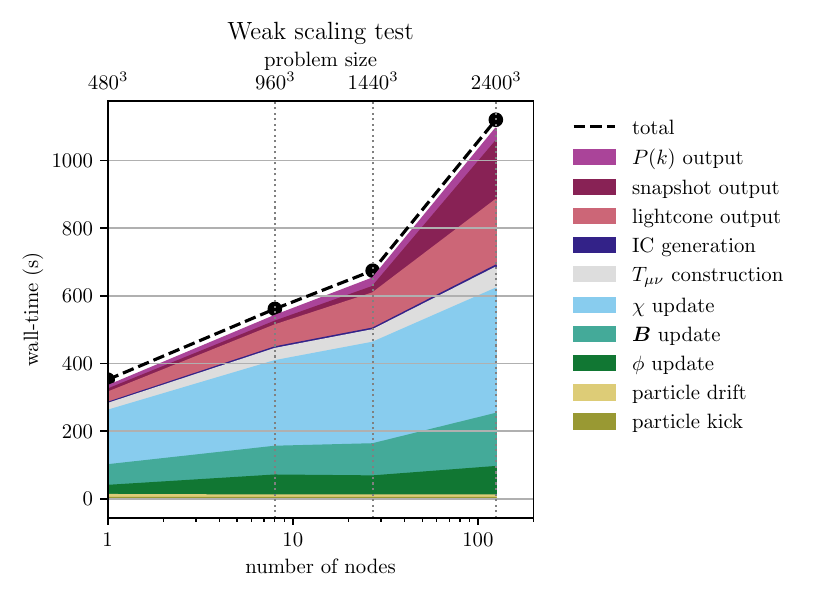}
    \caption{Results of a weak scaling test. The black dashed line indicates the overall performance as the problem size increases proportionately to the number of nodes. The different coloured bands show the relative contribution and the scaling behaviour of individual code components as indicated in the legend.}
    \label{fig:weakscaling}
\end{figure}

Weak scaling measures parallel efficiency as one increases the problem size and the number of processing elements proportionately, keeping the amount of work per processing element fixed up to overheads. This is a very useful metric because it indicates how the code performs in the limit of a very large problem size where using distributed memory at a massive scale is no longer optional. Fig.~\ref{fig:weakscaling} shows the performance of \texttt{gevolution} \texttt{2.0} for a weak scaling test using a simulation with $649$ integration steps, increasing the problem size (number of mesh cells and number of particles) proportional to the number of nodes. In this test, a constant wall-time would indicate perfect scaling. We observe near-perfect scaling for the particle updates and generation of initial data, but these components make up only a small fraction of the cost. The metric solvers, which rely heavily on MPI communication within the 3D FFTs, show good scaling overall, but the single-node case performs particularly well because the CUDA-aware MPI layer can communicate entirely on the intra-node P2P NVLink bypass. At very large problem size, I/O operations to the file system start to degrade the overall scaling performance. To some extent, this is again due to a saturation of the I/O throughput: for the largest run, which had $2400^3$ particles, we have to write more than $4\,\mathrm{TB}$ to the parallel file system, which takes a noticeable fraction of the $\sim 1120\,\mathrm{s}$ wall-clock time.

\subsection{Speed-up with respect to CPU-only version}

To measure the performance gain of the new GPU implementation over the previous version \texttt{1.3} of \texttt{gevolution}, we run \texttt{gevolution} \texttt{1.3} on the same hardware, but using CPUs only. Specifically, we run the simulation used for the strong-scaling analysis on three nodes with 96 MPI ranks per node. This means that we use only ${}^1\!/\!_3$ of the 288 CPU cores available on each node, a restriction made necessary due to the network configuration that would not allow a higher number of MPI endpoints per node. To account for this fact, we calculate the resource consumption (measured in node hours) consistently as
\begin{equation}
    (\text{wall-clock time}) \times (\#\,\text{nodes}) \times (\text{node utilisation})\,.
\end{equation}
To compare our CPU-only run to a similar run from the strong-scaling analysis, we pick the run performed on four nodes to have a mix of inter- and intra-node communication that is roughly comparable. In this case, we find a performance gain (reduction in total node hours consumed) of a factor of $\sim 4.5$ for the new GPU implementation.

Due to the significant differences in the computational backend, the outputs produced by the two versions of \texttt{gevolution} are not expected to be byte-identical. We verified that the differences between the results from \texttt{gevolution}\,\texttt{2.0} and \texttt{gevolution}\,\texttt{1.3} are consistent with noise from floating-point arithmetic and therefore well below the level of other numerical uncertainties.

\subsection{Comparison to other codes}

In our final performance analysis, we run a side-by-side comparison with two other $N$-body codes that have a significant footprint in the community, \texttt{Gadget-4} \citep{Springel:2020plp} and \texttt{PKDGRAV3} \citep{Potter:2016ttn}. Although the three codes have been developed for somewhat different purposes and therefore have different capabilities in terms of physical modelling, we choose a common baseline for this comparison: a pure Newtonian gravity $N$-body simulation. Specifically, we use \texttt{monofonIC}\footnote{\url{https://github.com/cosmo-sims/monofonIC.git}} \citep{Michaux:2020yis} to generate initial conditions for $1024^3$ particles from third-order Lagrangian perturbation theory at redshift $z_\mathrm{in} = 31$. The particles are then evolved independently with the three different codes up to redshift $z = 0$, each code producing a fixed set of snapshots that can be compared. The generation of initial data is therefore not part of the comparison, and this approach ensures that all three codes are solving exactly the same problem. We use a $\Lambda$CDM cosmology with $\Omega_\mathrm{m} = 0.3175$, $h = 0.6711$, and neglect the contribution of photons and massless neutrinos below $z_\mathrm{in}$. The simulation box was set to $1024\,h^{-1}\,\mathrm{Mpc}$.

Computationally, the three codes use different methods to solve the Newtonian gravitational dynamics. The comparison illustrates the trade-off between speed and small-scale accuracy, as well as the general benefits of hardware acceleration. In \texttt{gevolution}, a pure particle-mesh scheme with an FFT-based solver for the gravitational potential is used. Moreover, the mesh is kept uniform, which implies a fixed force resolution. Specifically, we choose a $2048^3$ mesh and use second-order gradients for force interpolation. \texttt{PKDGRAV3}, on the other hand, is a pure tree code that uses the fast multipole method to speed up the two-body force calculations. The advantage of using a tree algorithm is that the force resolution is adaptive and, therefore, much better suited to resolve the internal structure of collapsed halos. The fast multipole method is also designed to asymptotically outperform FFT-based methods for large problem sizes. Like \texttt{gevolution}\,\texttt{2.0}, \texttt{PKDGRAV3} is highly optimised for GPU architectures. Although \texttt{Gadget-4} can also be run as a pure tree algorithm, we run it in the default configuration of a mixed tree-particle-mesh scheme, where long-range forces are computed efficiently with the FFT method, while the tree algorithm adds short-range forces with adaptive force resolution. Unlike the other two codes, \texttt{Gadget-4} lacks a GPU implementation and runs exclusively on the CPU side.

\begin{table}
	\centering
    \begin{tabular}{cccccc}
		\toprule
		 & total & disk I/O & solve $\phi$ & kick/drift & CIC projection\\
		\midrule
		time (s) & 487 & 6.1/5.3 & 446 & 3.0/18.3 & 3.2\\
		\midrule
        node h & 0.54 &  \multicolumn{4}{l}{$\leftarrow$ \texttt{gevolution} \texttt{2.0}} \\
        \cline{1-2}
	\end{tabular}
	\caption{Performance profile of a Newtonian run with $1024^3$ particles and a $2048^3$ mesh. The run was executed on 4 nodes using 16 MPI ranks. Disk I/O consists of reading initial data (I) and writing five particle snapshots (O) to the parallel file system. The simulation ran for 434 integration steps, each consisting of one cloud-in-cell (CIC) projection to construct the density field, one computation of the gravitational potential $\phi$ using 3D fast Fourier transforms, one kick step to update the particle momenta, and one drift step to update the particle positions.}
	\label{tab:benchmark}
\end{table}

\begin{table}
    \centering
    \begin{tabular}{ccccccc}
		\toprule
		 & total & disk I/O & gravity & drift & tree & domain \\
		\midrule
		time (s) & 10183 & 156/66.3 & 9032 & 60.0 & 444 & 425\\
		\midrule
        node h & 11.3 &  \multicolumn{5}{l}{$\leftarrow$ \texttt{PKDGRAV3}} \\
        \cline{1-2}
	\end{tabular}
	\caption{Performance profile of the Newtonian run with $1024^3$ particles using \texttt{PKDGRAV3}. The run was executed on 4 nodes using 16 MPI ranks.}
	\label{tab:benchmarkpkd}
\end{table}

\begin{table}
    \centering
    \begin{tabular}{cccccc}
		\toprule
		 & total & tree & PM & kick/drift & domain \\
		\midrule
		time (s) & 342175 & 185242 & 10418 & 2419 & 124505\\
		\midrule
        node h & 95.0 &  \multicolumn{4}{l}{$\leftarrow$ \texttt{Gadget-4}}\\
        \cline{1-2}
	\end{tabular}
	\caption{Performance profile of the Newtonian run with $1024^3$ particles using \texttt{Gadget-4}. The run was executed on 3 nodes using 288 MPI ranks. This uses only $^1\!/\!_3$ of the available CPU cores on each node due to restrictions on the number of MPI endpoints per node in the network configuration. The calculation of node hours includes this factor of $^1\!/\!_3$ for the partial utilisation of nodes.}
	\label{tab:benchmarkgadget}
\end{table}

We run all three codes without any specific performance tuning. For \texttt{Gadget-4}, we set \texttt{TypeOfOpeningCriterion=1}, \texttt{PMGRID=2048}, \texttt{ASMTH=3.0}, \texttt{RCUT=6.0}, \texttt{ErrTolTheta=0.5}, \texttt{ErrTolForceAcc=0.005}, \texttt{ErrTolIntAccuracy=0.025} and \texttt{MaxSizeTimestep=0.005}, following \citet{Bairagi:2025sux}. The comoving softening scale was set to $25\,h^{-1}\,\mathrm{kpc}$. For \texttt{PKDGRAV3}, we use $217$ scheduled global time steps, \texttt{bEpsAccStep=True}, and leave accuracy settings at defaults, corresponding to \texttt{dEta=0.2} and a comoving softening scale of $20\,h^{-1}\,\mathrm{kpc}$. For \texttt{gevolution}, we set \texttt{Courant}\,\texttt{factor=36} and \texttt{time}\,\texttt{step}\,\texttt{limit=0.04}, resulting in $434$ time steps. The parameter files for the runs are available in the repository linked below.

Table~\ref{tab:benchmark} shows the performance of the new GPU implementation of \texttt{gevolution} for the Newtonian simulation. As might be expected, the execution time is completely dominated by the FFT-based solver for the gravitational potential $\phi$, as all other operations are essentially local and require little to no communication between the MPI ranks.

Table~\ref{tab:benchmarkpkd} shows the performance of \texttt{PKDGRAV3} for the same problem. Again, the execution time is dominated by the gravitational force calculation, for similar reasons as before. Compared to \texttt{gevolution}\,\texttt{2.0}, \texttt{PKDGRAV3} runs approximately $20$ times slower, which can probably be explained by the added cost of adaptive force resolution and the corresponding adaptive time stepping. In other words, the advantage in speed seen in \texttt{gevolution} comes, at least partly, at the cost of loss of accuracy when it comes to resolving the smallest scales. The statement which code `performs better' therefore depends on the scientific question one wants to study. We also stress that the timing ratios that we report can depend on the hardware (including the network architecture) and on specific aspects of the benchmark problem.

Lastly, Table~\ref{tab:benchmarkgadget} shows the performance of \texttt{Gadget-4} for the same problem. This code does not utilise GPUs and therefore runs yet an order of magnitude slower than \texttt{PKDGRAV3} on the same hardware. However, this code may still be a preferred choice on systems where no GPUs are available or in scientific studies that rely on some of the unique capabilities of \texttt{Gadget-4} to model non-gravitational physics. As a general point, one should always keep in mind that the three codes have been developed with different objectives in mind, and each of them has some features that make it uniquely suited for one specific purpose or another. But the general take-away of this comparison is that the relativistic framework implemented in \texttt{gevolution} is very much competitive with the state of the art in $N$-body simulations.

\begin{figure}
	\includegraphics[width=\columnwidth]{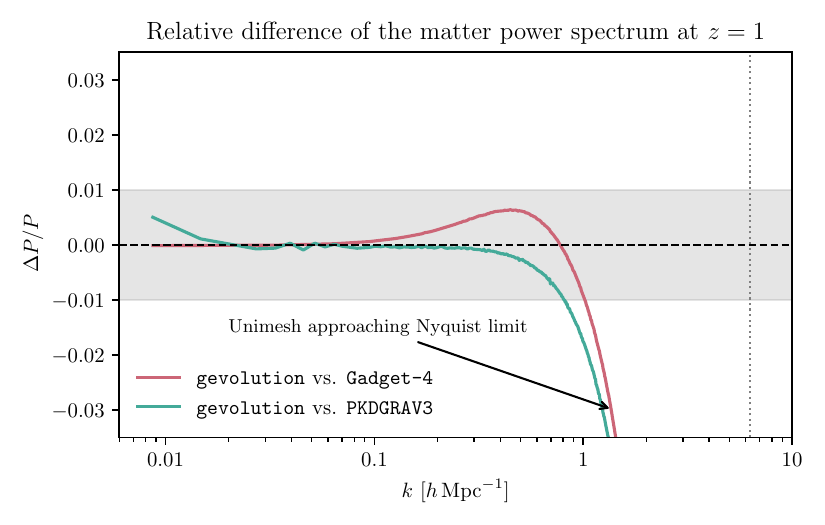}
    \includegraphics[width=\columnwidth]{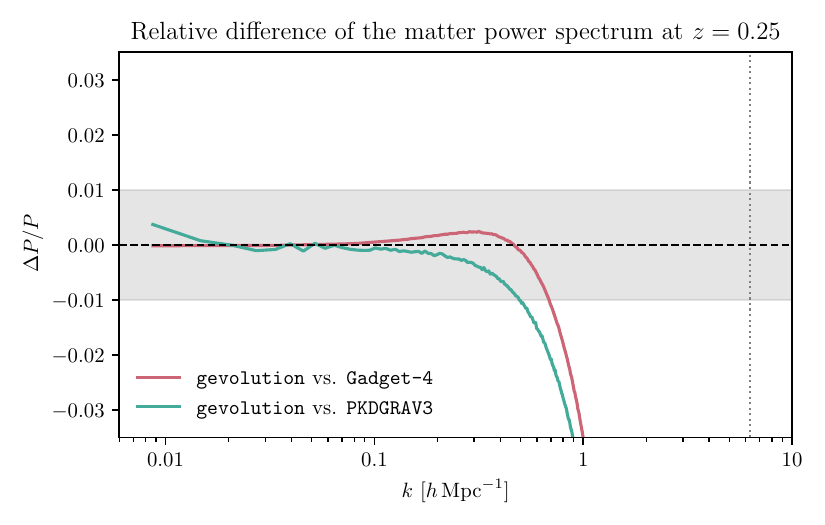}
    \caption{Comparison of matter power spectra for Newtonian simulations run with different codes. The top panel refers to redshift $z=1$ and the bottom panel to redshift $z=0.25$. The rose-coloured lines show the relative difference between \texttt{gevolution} and \texttt{Gadget-4}, whereas the teal-coloured lines show the relative difference between \texttt{gevolution} and \texttt{PKDGRAV3}. Positive numbers mean that the result from \texttt{gevolution} is larger than the respective reference. The grey bands indicate the $\pm 1\%$ region and the vertical dotted line marks the Nyquist limit of the mesh used in the \texttt{gevolution} run.}
    \label{fig:pk}
\end{figure}

To be more quantitative, Fig.~\ref{fig:pk} shows the relative differences between the matter power spectra of the simulation done with \texttt{gevolution} and those done with \texttt{PKDGRAV3} and \texttt{Gadget-4}, respectively. The top panel shows the results for redshift $z=1$ and the bottom panel shows redshift $z=0.25$. At the largest scales, \texttt{gevolution} and \texttt{Gadget-4} agree almost perfectly, while \texttt{PKDGRAV3} shows a slight, sub-percent deficit in power that we attribute to the behaviour of the tree algorithm relative to the particle-mesh approach shared between the former two codes. However, on intermediate scales, \texttt{PKDGRAV3} and \texttt{gevolution} are astonishingly consistent, at roughly one part per thousand, while \texttt{Gadget-4} deviates by up to almost $1\%$. Comparing to a previous \texttt{Gadget-4} benchmark used in the preprint version of this manuscript, we find that this mismatch depends on parameter tuning and can probably be mitigated further. At short scale, the lack of adaptive force resolution in \texttt{gevolution} becomes apparent through a suppression of power as one approaches the Nyquist limit of the uniform mesh, indicated as vertical dotted line in the plots. For our choice of mesh, the power spectrum remains within $1\%$ of the \texttt{Gadget-4} reference up to $k \sim 1\,h\,\mathrm{Mpc}^{-1}$ for redshift $z=1$ and up to $k \sim 0.7\,h\,\mathrm{Mpc}^{-1}$ for redshift $z=0.25$. Therefore, if the stated accuracy is sufficient for the scientific question at hand, the new GPU implementation of \texttt{gevolution} offers a very performant alternative to tree-based $N$-body methods.

\section{Conclusions}
\label{sec:conclusions}

With the release of version \texttt{2.0} of \texttt{gevolution}, we provide the community with a fast and scalable computational tool for relativistic $N$-body simulations in cosmology. The new implementation takes advantage of the widespread availability of GPU hardware in current HPC architectures. Our numerical framework is based on a pure particle-mesh algorithm with fixed resolution, keeping the mapping between configuration space and Fourier space straightforward. It allows for the consistent treatment of arbitrary relativistic sources of stress-energy, which has interesting applications in the context of dynamical dark energy and other cosmological scenarios. The move towards a GPU implementation may pay off particularly well for models whose equations can be formulated using a field language, and whose dynamics can then be solved numerically on a discrete mesh. It is our hope that the capability to include such dynamics in the nonlinear forward modelling of cosmological observables will help us to ultimately understand the true nature of the dark sector in our Universe.

\section*{Acknowledgements}

{\small The authors thank F.\ Hassani, F.\ Lepori, L.\ Mosimann (Nvidia) and I.\ Magkanaris (CSCS) for valuable contributions during code development and D.\ Potter for practical advice concerning \texttt{PKDGRAV3}. This work was supported by two grants from the Swiss National Supercomputing Centre (CSCS) under project IDs sm97 and lp165. This work was completed in part at EuroHack24 with CSCS, part of the Open Hackathons program. The authors would like to acknowledge OpenACC-Standard.org for their support. J.A. acknowledges financial support from the Swiss National Science Foundation and the Dr.\ Tomalla Foundation for Gravity Research. \O{}.C. acknowledges support from the European Union and the Czech Ministry of Education, Youth and Sports (Project No.\ FORTE – CZ.02.01.01/00/22\_008/0004632 and the e-INFRA CZ \text{PiD}:FTA-25-52).}

\section*{Code Release}

{\small The public release of \texttt{gevolution}\,\texttt{2.0} can be found at \url{https://github.com/gevolution-code/gevolution-2.0}. We also release a GPU-ready version of the \texttt{LATfield2} library which provides the parallelisation backend, available at \url{https://github.com/gevolution-code/LATfield2}.}

\section*{Data Availability}

{\small The configuration files and profiler reports of our benchmark runs can be found at \url{https://doi.org/10.5281/zenodo.24371348}.}


\bibliographystyle{mnras}
\bibliography{biblio}

\end{document}